\documentstyle[12pt]{article}

\newcommand{\e}{\epsilon}
\newcommand{\bb}{\begin{equation}}
\newcommand{\ee}{\end{equation}}
\newcommand{\p}{\partial}
\begin{document}

\begin{titlepage}

\begin{flushright}

BRX TH-405\\
ULB-TH-97/03\\

\end{flushright}

\begin{center}
{\large\bf Duality, Self-Duality, Sources and 
Charge Quantization in Abelian $N$-Form Theories}

\end{center}
\vfill

\begin{center}
{\large
S. Deser$^{a}$,
A. Gomberoff$^{b}$,
M. Henneaux$^{b,c}$ \\ and
C. Teitelboim$^{b,d}$}
\end{center}
\vfill

\begin{center}{\sl
$^a$ Department of Physics, Brandeis University,\\
Waltham, MA 02254, U.S.A.\\[1.5ex]

$^b$ Centro de Estudios Cient\'\i ficos de Santiago,\\
Casilla 16443, Santiago 9, Chile\\[1.5ex]

$^c$ Facult\'e des Sciences, Universit\'e Libre de
Bruxelles,\\
Campus Plaine C.P. 231, B--1050 Bruxelles, Belgium\\[1.5ex]

$^d$ School of Natural Sciences, Institute for 
Advanced Study,\\
Princeton, New Jersey 08540, U.S.A.

}\end{center}
\vfill

\begin{abstract}
We investigate duality properties of $N$-form
fields, provide a symmetric way of coupling them
to electric/magnetic sources, and check that these
charges obey the appropriate quantization requirements.  
First, we contrast
the $D=4k$ case, in which duality is a well-defined
SO(2) rotation generated by a Chern--Simons form
leaving the action invariant, and $D=4k+2$ where
the corresponding ostensibly SO(1,1) rotation is
not only not an invariance but does not even have
a generator.
When charged sources are included we show explicitly
in the Maxwell case how the usual Dirac quantization
arises in a fully symmetric approach attaching
strings to both types of charges.  Finally, for
$D=4k+2$ systems, we show how charges can be
introduced for self-dual $(2k)$-forms, and obtain
the $D=4k$ models with sources by dimensional reduction, tracing
their duality invariance to a partial invariance in
the higher dimensions.

\end{abstract}

\end{titlepage}

The ubiquitous current relevance of duality \cite{Olive} 
prompts
the present investigation, in the simplest context
of abelian $N$-form fields.  First, we establish a
fundamental difference between even and odd $N$-form
 systems; electric-magnetic duality is only 
definable for $(2k+1)$-form potentials, whose 
(free-field) actions are invariant under this SO(2)
rotation, while the ostensibly hyperbolic SO(1,1)
transformation for $2k$-forms is not even 
implementable as a canonical transformation, 
let alone an invariance. This difference
is traceable to the (non-)existence of Chern--Simons terms
as generators of duality transformations.  We next
couple electric/magnetic monopole sources in the 
specific context of Maxwell theory in Hamiltonian
formulation, using a Dirac string approach for both
types of charges.  We deduce within this context the
usual Dirac quantization condition in its manifestly
duality invariant form, despite the
seeming appearance of the infamous but (spurious) 
factor 2.  Finally, we return to $D=4k+2$ in the 
context of self-duality (which in turn is 
unavailable in $D=4k$).
Here we complete previous formulations by introducing
coupling to sources.  We also show how dimensional
reduction to $D=4k$ yields the manifestly 
duality-invariant action with sources
there and how a partial invariance in the higher
space underlies duality invariance in $D=4k$.

The field strengths are 2$k$ or $(2k + 1)$-forms 
which we uniformly denote
by $F$.  For simplicity, we work
in flat space but everything carries over to
arbitrary backgrounds since there are no covariant
derivatives.  The basic identities governing the dual
operation $^*\!F = \e F$ are
\bb
^{**}\! F = +F, \;\;\; D = 4k+2 ; \;\;\; 
^{**}\! F = -F, \;\;\; D = 4k \; .
\label{dual}
\ee

In the absence of sources, the equations of motion are
\bb
dF = 0, \; \; d^*\!F = 0,
\label{EOM1}
\ee
and are invariant under linear transformations of $F$ and
$^*\!F$,
\begin{equation}
\left( \begin{array}{c}
F^\prime\\ ^*\!F^\prime
\end{array}
\right)
=
B \left( \begin{array}{c}
F\\ ^*\!F
\end{array}
\right)
\equiv
\left( \begin{array}{ll}
a & b\\
c & d 
\end{array}
\right)
\left( \begin{array}{c}
F\\ ^*\!F
\end{array}
\right), \; \; \det B \not= 0.  \label{linear}
\end{equation}
In view of (\ref{dual}),
the condition that $^*\!F$ is the dual of $F$ 
imposes both $a=d$
and $c=b$ ($D= 4k + 2$) or $c=-b$ ($D= 4k$).
In the first case, the group defined by (\ref{linear}) is 
$R^+ \times Z_2 \times SO(1,1)$ since any 
invertible $2\times 2$ matrix
with $a=d$ and $c=b$ can be uniquely decomposed as
the product of (commuting) factors
\begin{equation}
B \equiv
\left( \begin{array}{ll}
a & b\\
b & a 
\end{array}
\right)
=
\left( \begin{array}{ll}
\lambda & 0\\
0 &  \lambda 
\end{array}
\right)
\left( \begin{array}{ll}
0 & 1\\
1 & 0 
\end{array}
\right)^\epsilon
\left( \begin{array}{ll}
 \pm\cosh \alpha & \pm\sinh \alpha\\
\pm\sinh \alpha & \pm\cosh \alpha 
\end{array}
\right)
\label{decomp1}
\ee
for some appropriate $\alpha$.  Here, $\lambda
= (\vert \det B \vert)^{\frac{1}{2}}$ and 
$\epsilon = 0$ ($\det B > 0$) or $\epsilon = 1$
($\det B < 0$).
In the second case, the group defined by (\ref{linear})
and $a=d$, $c=-b$ is $R^+ \times SO(2)$ since one has
\begin{equation}
B \equiv
\left( \begin{array}{ll}
a & b\\
-b & a 
\end{array}
\right)
=
\left( \begin{array}{ll}
\lambda & 0\\
0 &  \lambda 
\end{array}
\right)
\left( \begin{array}{ll}
\cos \alpha & \sin \alpha\\
- \sin \alpha & \cos \alpha 
\end{array}
\right)
\label{decomp2}
\ee
for  $\lambda
=  (\det B)^{\frac{1}{2}}$  and some appropriate $\alpha$.

The above groups are symmetry groups of the equations of 
motion, with $F$ regarded as the independent variable.  
However, in order to be  true symmetries of the theory, 
the duality transformations
should leave the action invariant.  Since the action 
principle involves the potential $A$ as basic field, 
$F= dA$ being a derived quantity, 
this means that one
must rewrite the duality transformations in terms 
of the potential, and 
verify whether they leave the free field action
\bb
I = \int (dx) tr [FF-^*\! F ^*\! F] 
\label{freeaction0}
\ee
invariant.  [We have dropped an irrelevant
multiplicative numerical factor and have used 
(\ref{dual}) as well as $tr A ^*\!B =
tr ^*\!A B$, where the trace is on the spacetime indices
to write tr$(FF + ^*\!F^*\!F)=0$ in {\it all}
dimensions.]

Once the potential is introduced, the equation $dF=0$ 
becomes an identity,
while $d ^*\!F =0$ remains  an equation of motion.
Thus, one cannot rotate $F$ into $^*\!F$ off-shell, 
and the best that can
be achieved is to define the duality transformation 
of the potential in such
a way that the induced transformation of the field 
strengths reduces to (\ref{linear}) 
on-shell, with $B$ given by (\ref{decomp1}) 
or (\ref{decomp2}).
This task is carried out explicitly below, but
the results can be anticipated more quickly by 
considering the energy-momentum tensor.
Because the duality transformations
must commute with the Lorentz transformations
if they are to be symmetries of the action, they
should leave the energy-momentum tensor
(whose moments generate the Poincar\'e group) on-shell 
invariant.  [This condition is of course also
required for  invariance of the gravitational coupling.]

The stress
tensor may be written uniformly in all dimensions as
\bb
2T^{\mu\nu}= F^{\mu} F^{\nu} + \, ^{*} \! F^{\mu}\, 
^{*}\! F^{\nu} \;,
\label{energymomentum}
\ee
where the unwritten indices are summed over.
Conformal invariance of these theories is reflected
in the identical tracelessness of $T^{\mu\nu}$ in
all dimensions, remembering that tr$(F^2 + \, ^*\!F^2)
\equiv 0$.  Since the $FF$-term and the 
$^{*} \!F ^{*}\!F$-term have the same
sign, as required by positive-definiteness
of $T_{00}$, it is clear that the energy-momentum tensor is
only invariant under $O(2)$ rotations of $F$ into $^{*}\!F$
and is {\em not} invariant under hyperbolic rotations.
The subgroups of the above groups that leave
$T_{\mu \nu}$ invariant are thus the intersections of
(\ref{decomp1}) or (\ref{decomp2}) with $O(2)$ and
are equal to the factors
\bb
G = Z_2 \; , \hbox{ for } D = 4k + 2
\ee
and
\bb
G = SO(2) \; , \hbox{ for } D = 4k \, .
\ee
We shall call these groups the ``duality groups".  There is 
therefore an essential difference between the cases 
$D = 4k + 2$ and $D = 4k$.  While the duality group 
is a one-parameter continuous
group in the latter, it is a discrete group with just two
elements in the former.\footnote{Note that in the
$D = 4k + 2$ case, the $SO(1,1)$ transformation given by 
minus the unit 
matrix leaves also the stress tensor invariant.  So,
the duality group is actually $Z_2 \times Z_2$. However, 
this additional $Z_2$ is rather trivial
since it amounts to changing the sign of $A$, $F$ 
and $^{*}\!F$.
For this reason, it has not been mentioned in the text.}

We now verify that these conclusions are correct and
that $SO(2)$-duality rotations do indeed leave the action
invariant for $D=4k$, while hyperbolic
rotations are not symmetries of the action for $D=4k +2$.
It would be meaningless to transform the $F$'s
formally in the action, since they are not the
dynamical variables.
As stressed above, in order to discuss any putative symmetry
transformation,
one must first be able to implement it on the potential $A$.
To this end, the Hamiltonian
formulation is most convenient; while not manifestly
Lorentz invariant, it does preserve manifest gauge
invariance.  The required discussion was first given
long ago \cite{DHT,Des} for Maxwell theory
in four dimensions, which is illustrative of the general 
$D=4k$ case.  

In the absence of sources, the canonical
variables are the transverse, gauge invariant pair
$({\bf E}_T, \; {\bf A}_T )$.  It is convenient to write
${\bf E}_T = \mbox{\boldmath $\nabla$} \times {\bf Z}_T$ 
(to parallel
${\bf B} = \mbox{\boldmath $\nabla$} \times {\bf A}_T)$ 
and then
define the 2-vector ${\bf A}_a = ({\bf A} , {\bf Z} )$
as well as the derived quantities
\bb 
{\bf B}^a = \mbox{\boldmath $\nabla$} \times {\bf A}^a \; ,
\;\;\; {\bf E}^a = \dot{\bf A}^a \; ,
\ee
dropping the ``$T$" notation hereafter.
In terms of these, the manifestly $SO(2)$ invariant 
canonical action becomes\footnote{The action (\ref{free1})
was written in \cite{DHT} starting directly from the Maxwell
action in Hamiltonian form and solving the Gauss law.  It
was rederived independently in \cite{SSen}. A
manifestly covariant formulation was given recently
in \cite{Berk}, but it requires an infinite 
number of fields.}
\bb
I = \frac{1}{2} \int d^4 x (\e_{ab} {\bf B}^a \cdot
{\bf E}^b - {\bf B}^a \cdot {\bf B}^b \delta_{ab})\;.
\label{free1}
\ee
Note the absence of the Lagrange multipliers $A_0^a$
(as well of course as of the longitudinal gauge
components) since the corresponding Gauss constraints
$\mbox{\boldmath $\nabla$} \cdot {\bf E} = 
\mbox{\boldmath $\nabla$} \cdot {\bf B}=0$ 
are already incorporated.  
The invariance of (\ref{free1}) under $SO(2)$ rotations
\begin{equation}
\left( \begin{array}{c}
{\bf A}^\prime_1\\{\bf A}^\prime_2
\end{array}
\right)
= 
\left( \begin{array}{ll}
\cos \alpha & \sin \alpha\\
-\sin \alpha & \cos \alpha
\end{array}
\right)
\left( \begin{array}{c}
{\bf A}_1\\{\bf A}_2
\end{array}
\right) \label{dualdual}
\end{equation}
is obvious since both $\epsilon_{ab}$ and
$\delta_{ab}$ are invariant tensors
for $SO(2)$.  The transformation (\ref{dualdual})
rotates the $B$'s  (and the $E$'s) among themselves and
reduces to (\ref{linear}) on-shell since
$\epsilon_{ab} {\bf B}^b$ and ${\bf E}^a$ coincide there.

The invariance of the kinetic term implies that the
transformation (\ref{dualdual}) is a canonical one.
The generator that performs
the rotations is simply the Chern--Simons term
\bb
G = - \frac{1}{2} \int d^3x \, {\bf A}^a \cdot
{\bf B}^b \delta_{ab} \; .
\label{ChernSimons}
\ee
The invariance of the Hamiltonian is equivalent
to $[H, G]= 0$. 
The duality invariance of
the stress tensor is likewise obvious, as it depends
on either ${\bf B}^a {\bf B} ^a$ or $\e_{ab} 
{\bf B}^a {\bf B}^b$.  

By contrast, the zero-form (mod 2$k$) potential has
canonical action
\bb
I = \int d^2x \left[ E\dot A - \frac{E^2 + 
A^{\prime 2}}{2} \right]
\label{odd}
\ee
which seems formally very like its Maxwell analog
above.  The $SO(1,1)$ transformations read, in terms of the dynamical 
variables $A$ and $E$ (or equivalently, $A$ and $Z$
with $Z^\prime = E$)
\begin{equation}
\left( \begin{array}{c}
 A_{NEW}\\ Z_{NEW}
\end{array}
\right)
=
\left( \begin{array}{ll}
\cosh \alpha & \sinh \alpha\\
\sinh \alpha & \cosh \alpha
\end{array}
\right)
\left( \begin{array}{c}
 A\\ Z
\end{array}
\right) \label{dualwrong}
\end{equation}
and are quite similar to (\ref{dualdual}).
However, they do not leave the symplectic (kinetic) term
invariant and hence do not define canonical transformations.
Indeed, the infinitesimal transformation $\delta A = Z$, 
$\delta Z = A$ yields $\delta(E \dot A) = A^\prime \dot A
+ Z^\prime \dot Z$, which is {\em not} a total derivative.
The same conclusion would be reached for $SO(2)$ rotations
of $A$ and $Z$, since changing the sign of the $Z^\prime \dot Z$
term in  $\delta(E \dot A)$ does not
help.  The absence of a canonical generator
analogous to (\ref{ChernSimons}) for the duality rotations 
in $4k + 2$ dimensions is of course also clear from the
fact that there are no Chern--Simons terms
in 1 (mod 4$k$) space dimensions for the 2$k$-form
spatial component potentials. Note furthermore  that 
$SO(1,1)$ does not even leave the
Hamiltonian invariant. We conclude that duality 
rotations---whether SO(1,1) or even SO(2)-- are not
symmetries of the theory and cannot be implemented as
canonical transformations.  
The $Z_2$ transformation of (\ref{decomp1})
is however a canonical transformation that leaves the 
Hamiltonian 
invariant since it exchanges $E$ with $A'$; the generating
function for that transformation is just 
$\int dx A^\prime_{OLD}
A_{NEW}$.  One may actually decompose the fields according
to the two irreducible representations of $Z_2$
(chiral and anti-chiral $(2k)$-forms) and write actions
for the individual chiral and anti-chiral 
components \cite{FloJa,HT2}.\footnote{One reaches similar
conclusions in a spacetime
with Euclidean signature.  The signs in (\ref{dual})
are reversed, and the relative sign in the
energy-momentum tensor is also changed.  The duality group in
$4k$ dimensions is now $SO(1,1)$ and leaves the action invariant.
By contrast, duality (whether $SO(2)$ or $SO(1,1)$) does
not leave the action invariant in $4k + 2$ dimensions.
Only $Z_2$ is a symmetry.  One may construct actions
for chiral and anti-chiral $2k$-forms in $4k +2$ dimensions,
but the chirality condition involves a factor of $i$.  The
corresponding actions are just the Euclidean continuations
of the (first-order) actions of  \cite{FloJa,HT2}.}

We now turn to the introduction of electric and
magnetic sources of the Maxwell field, using Dirac
strings \cite{Dirac} symmetrically to describe both, 
rather than
using Coulomb fields for one type and strings for
the other.  Call particle $A$'s charge and mass
$(q_A , m_A)$ and attach a string $y_A(\sigma_A,
\tau_A )$ to it, the range of the spacelike 
$\sigma_A$ being $(0,\infty )$, and $\tau_A $ the
proper time; a particle's trajectory is specified
by $z^\mu_A (\tau_A ) = y^\mu (\sigma_A = 0, \;
\tau_A )$.  Instead of Coulomb fields, we add to the 
free fields $({\bf E}^a ,{\bf B}^a )$ the following
string worldsheet terms:
\begin{eqnarray}
{\bf B}^a &=& \mbox{\boldmath $\nabla$} \times {\bf A}^a + \sum_A
\int dy^0_A \wedge d{\bf y}_A \; q^a_A \: \delta^4
(x-y_A), \label{B}\\ 
{\bf E}^a &=& \dot{\bf A}^a + \frac{1}{2} \:
\sum_A \int d {\bf y}_A \wedge \times d {\bf y}_A \:
q^a_A \: \delta^4 (x-y_A) 
\label{E}
\end{eqnarray}
where in (\ref{E}) there is also a vector cross
product indicated.  Note that the symmetric Gauss
law
\bb
\mbox{\boldmath $\nabla$} \cdot {\bf B}^a
 = \sum  q_A^a
\delta^{(3)}(x - z_ A) 
\label{gausslaw}
\ee
is implied by (\ref{B}).
The total action is then the sum of the Maxwell 
action (\ref{free1}) with the redefined fields of 
(\ref{B})-(\ref{E})
together with the current interaction and free
particle contributions,
\begin{eqnarray}
\lefteqn{I_{ist} = I_{max} + I_{int.}+I_p =} 
\nonumber \\
&& \frac{1}{2} \int d^4x ({\bf B}^a \cdot 
{\bf E}^b \e_{ab}
- {\bf B}^a \cdot {\bf B}^a ) + \frac{1}{2} 
\sum_A \epsilon_{ab}
q^b_A \int {\bf A}^a (z_A) \cdot d {\bf z}_A
\nonumber \\
& - & \sum_A \: m_A \int \sqrt{-(dz^\mu )^2} \; .
\label{actionsources1}
\end{eqnarray}
Varying $I_{tot}$ with respect to ${\bf A}^a$ gives
\bb
{\bf B}^a + \e_{ab} \mbox{\boldmath $\nabla$} 
\times {\bf B}^b +
\sum_A q^a_A \int \delta^4 (x -z_A) d{\bf z} = 0
\label{B2}
\ee
or equivalently, $\mbox{\boldmath $\nabla$} \times (\e_{ab}
{\bf B}^b + {\bf E}_a ) = 0$, which together with
(\ref{gausslaw}) are the Maxwell equations with 
magnetic/electric
fields/charges described by $({\bf B}^1, q^1_A)$, 
$({\bf B}^2 , q^2_A)$.  The particle equations from
varying $z^\mu_A$ yield the correct Lorentz force 
law,
\begin{eqnarray}
m \ddot{\bf z}_A & = & q^a_A {\bf B}^a 
(z_A) \dot z^0_A
+ \e_{ab} q^a_A \, {\bf B}^b (z_A) \times 
\dot{\bf z}_A
\nonumber \\
m \ddot z^0_A & = & q^a_A {\bf B}^a \cdot 
\dot{\bf z}
\label{Lorentz1}
\end{eqnarray}
where dots denote derivatives with respect to proper
time.  There remain the string coordinates, to be
varied (leaving the history of the corresponding pole
fixed).  Properly, this does not lead to any new
equations (on Maxwell shell) but consistency 
requires that no pole of charge $q^a_A$ cross a
string attached to any other particle $B$. Note  
that, in the absence of dyons,  {\it i.e.}, when
there are only pure electric and pure magnetic 
poles---say one of each---our action 
(\ref{actionsources1}) reduces
to the original Dirac form \cite{Dirac} up to a total
 divergence
in field space which enables one
to drop the electric string; they are therefore equivalent.
Finally, note the unusual $\frac{1}{2}$ factor in 
the ``${\bf A} \cdot{\bf j}$" term of 
(\ref{actionsources1}); as we have
seen, the correct Lorentz force law nevertheless
comes from it together with the string-related
contributions implicit in (\ref{B})-(\ref{E}), which
bring in another  factor of
$\frac{1}{2}$.

We are now in a position to derive the usual
charge quantization condition in its manifestly 
duality-invariant form
\cite{Dirac,Schwinger,Zwan}
\bb
\epsilon_{ab} \bar q^a q^b = nh \; .
\label{quanti1}
\ee
Its validity is most directly noted from the Lorentz
force law (\ref{Lorentz1}) which involves the U(1) 
connections ${\bf A}^a$, {\it not} $\frac{1}{2} \, 
{\bf A}^a$, then following the usual Wu--Yang arguments 
\cite{Wu}. However, it behooves us to show that even 
at the level of the action (\ref{actionsources1}) 
the $\frac{1}{2}$ does {\it not} mean that a 
doubling of the right side of (\ref{quanti1}) is required.
The apparent paradox here is similar to the 
complications as to when a wave function's 
phase must change by $2\pi n$ in discussing
anyons in (2+1) dimensions.  Indeed, the wave
functional $\Psi$'s dependence on the string
coordinates $y_A$ implies that it acquires a
phase $\frac{1}{2\hbar}  \epsilon_{ab}
\bar q^aq^b$ as the string of the charge
$\bar q^a$, say, is passed around that of some
other $q^b$.  However, there is no requirement
that $\Psi$ be single-valued under this motion
because the configuration space is not simply
connected; a wave function is perfectly 
permitted to acquire a phase when the loop in
question is {\it not} (as is the case in
general here) contractible to a point.  However,
if one does a ``double pass" in which the string
attached to $\bar q^a$ is passed around another
$q^b$, while that of $q^b$ is passed around
$\bar q^a$, then that process {\it is} 
connected to the identity, so that the total
phase acquired must be a multiple of $2\pi$.
But this double pass is precisely {\it twice}
the above phase, $2 \times (\frac{1}{2\hbar}
\, \epsilon \bar qq)$.  In the original Dirac
formulation there is neither the $\frac{1}{2}$
factor to start with in the interaction, nor 
a multiply connected configuration space to
compensate for it, hence (\ref{quanti1}) always 
emerges in both representations, as it must.

Finally, we discuss some new results
concerning self-dual systems in $D = 4k+2$
of which the 2-form potential 
$A_{\Lambda\Delta}$ in $D=6$ is the first
interesting case beyond the $D$=2 chiral scalar
field.  Here we extend the original 
non-manifestly Lorentz covariant
formulation \cite{HT2} by introducing 
sources.\footnote{The
difficulty of writing a finite-component 
self-dual covariant form is simply that because
of the identity $F^2 =-^*\! F^2$, the (anti)
self duality condition $^*\! F =\pm F$ when
squared implies $^*\! F^2 = +F^2$ and thus
$F^2=0$.  Since $F ^*\! F$ is a total derivative, there
is no non trivial covariant expression that is quadratic
in the field strength.
Some attempts to bypass this difficulty by
introducing auxiliary fields -- in most cases an
infinite number of them -- are described in 
\cite{Berk,%
DH,Bengtsson,Pasti}.}
We will also reduce the
2-form action to Maxwell by dimensional
reduction and obtain directly the manifestly
duality invariant action (\ref{actionsources1}).  
In the process, we will trace the
sources of Maxwell duality invariance to a
particular rotation in the higher dimensional
space.

The field strengths
are defined as usual as
\bb
F_{\Gamma \Delta \Lambda} = \p_{\Gamma} A_{\Delta \Lambda}
+ \p_{\Delta} A_{\Lambda \Gamma} + \p_{\Lambda} 
A_{\Gamma \Delta}.
\label{fieldstrength}
\ee
The sourcefree action for a chiral 2-form is given
by the first-order expression \cite{HT2}
\bb
I= \frac{1}{4} \int d^6x [E^{AB} B_{AB} -
B^{AB} B_{AB}]
\label{freeaction2}
\ee
where the electric and magnetic components $E^{AB}$
and $B^{AB}$ are defined through 
\begin{eqnarray}
E^{AB} &=& - F^{0AB}, \\
B^{AB} &=& \frac{1}{6} \epsilon^{ABCDE} F_{CDE} .
\end{eqnarray}
Here, capital roman letters take the spatial values 
$(a, 4,5)$ with $a=1,2,3$.  Similarly, we introduce 
capital greek indices with values
$\Lambda = (\lambda, 4,5)$, $\lambda = 0,1,2,3$.  
The separation-out of two spatial directions ($4,5$) is 
motivated by
the reduction to four dimensions performed below.
The equations of motion are obtained by varying the action
with respect to the 2-form components $A_{AB}$ ($A_{0A}$
drops out) and are equivalent to the self-duality condition
\bb
E^{AB} = B^{AB}.
\label{chirality}
\ee

It is possible to introduce sources while
maintaining the chirality condition (\ref{chirality}), 
provided these sources have equal magnetic and electrical
charges.  That is, the electric and magnetic currents
must be equal,
\bb
J^{\Lambda \Delta}_{e} = J^{\Lambda \Delta}_{m} \equiv 
J^{\Lambda \Delta}.
\ee
The action describing the coupling is obtained from the
free action (\ref{freeaction2}) by (i) adding to it
the minimal coupling term $ A_{\Lambda \Delta} J^{\Lambda \Delta}$;
and (ii) redefining the fields $E^{AB}$
and $B^{AB}$ by including a contribution from the sources 
and such that $B^{AB}$ fulfills
\bb
\p_B B^{AB} \equiv J^{0A}
\ee
(and no longer $\p_B B^{AB}\equiv 0$) identically.  Explicitly,
\begin{eqnarray}
E^{AB} &=& - F^{0AB} + \frac{1}{6} \epsilon^{ABCDE} G_{CDE}, \\
B^{AB} &=& \frac{1}{6} \epsilon^{ABCDE} F_{CDE} - G^{0AB},
\end{eqnarray}
where $F_{\Gamma \Delta \Lambda}$ is still given by
(\ref{fieldstrength}) and where $G_{\Gamma \Delta \Lambda}$ 
has Dirac-string type singularities 
and is completely determined by the sources through
\bb
\p_{\Lambda} G^{\Gamma \Delta \Lambda} + J^{\Gamma \Delta} = 0
\ee
(see below).
So, the action is
\bb
I= \frac{1}{4} \int d^6x [E^{AB} B_{AB} -
B^{AB} B_{AB} - A_{\Lambda \Omega} J^{\Lambda \Omega}].
\label{actionsources2}
\ee
The temporal components  $A_{0 B}$ again drop
out and the equations of motion obtained by varying the spatial
components $A_{AB}$
are equivalent to the chirality condition (\ref{chirality}).

If one takes as source the elementary object to which the
2-form naturally couples, namely, a charged string, then 
$J^{\Lambda \Omega}$
is given by 
\bb
J^{\Lambda \Omega}(x) = e \int_{WS} \delta^{(6)}(x-z) 
dz^\Lambda 
\wedge dz^\Omega  
\ee
where the integral is taken along the string world sheet 
$z^{\Lambda}
(\tau, \sigma)$ and
where $e=g$ is the electric ($=$ magnetic) strength of the 
string.  The field
$G_{\Gamma \Delta \Lambda}$ has support on a membrane 
emanating from the string and is given by \cite{CT}
\bb
G^{\Gamma \Delta \Lambda}(x) =  e \int_{MWS} 
\delta^{(6)}(x-y)
dy^\Gamma \wedge dy^ \Delta \wedge dy^\Lambda.
\ee
Here, the integral is taken over the world sheet of the 
membrane $y^\Lambda
(\tau, \sigma, \rho)$ with $0 \leq \rho < \infty$ and
$y^\Lambda(\tau, \sigma, 0) = z^\Lambda(\tau, \sigma)$.
Varying the action, including the string's kinetic term,
with respect to the string coordinates yields the string 
equations of motion, with the appropriate Lorentz force.  
Varying the action with respect
to the membrane coordinates yield no equation, provided the 
membrane does not intersect any other string 
(``Dirac veto").  In the quantum regime, 
one finds that the string electric 
(= magnetic) strength $e$ fulfills the
quantization condition
\bb
e g \equiv e^2 =  n h.
\label{quanti2}
\ee
This condition can be derived either \`a la Wu-Yang \cite{Wu,HT}, 
or by requiring  as above that the membrane remain
quantum-mechanically unobservable \cite{CT,Nepo}.  It has 
been used recently in \cite{Perry}.  One striking feature of 
(\ref{quanti2}) is that it is {\it not} invariant under 
$SO(2)$ or $SO(1,1)$ rotations
of the electric and magnetic charges (the product $eg$ is not 
invariant). But this is all right since neither 
$SO(1,1)$ nor $SO(2)$ duality is a symmetry of the
action in six dimensions.  Only $Z_2$ is, and the quantization 
condition remains clearly unchanged if one interchanges
(the equal-valued) $e$ and $g$.

If the spacetime has the topology $R^4 \times T^2$, with two spatial
coordinates compactified on a torus $T^2$, one may naturally consider
sources of higher dimensions, here membranes
that are wrapped on the torus.
These extended objects thus have one more dimension than the
elementary source to which the 2-form couples and correspond, say, 
to charged rings in electromagnetism.  We take the coordinates
$x^4$ and $x^5$ to be along
the torus.  The spacetime history of the membrane is
$x^\mu = z^\mu (\tau)$, $x^4= x^4$, $x^5=x^5$ (we take
$\tau$, $x^4$ and $x^5$ as parameters).  The current does not depend
on the internal coordinates $x^4$, $x^5$ and is assumed to be
of the form
\begin{eqnarray}
J^{\lambda \mu} &=& 0, \\
J^{4 5}  &=& 0, \\
J^{4 \lambda} &=& g^4 \int_W \delta^{4}(x^\mu -z^ \mu) dz^\lambda, \\
J^{5\lambda} &=& g^5 \int_W \delta^{4}(x^\mu -z^ \mu) dz^\lambda,
\label{current}
\end{eqnarray}
where W is the trajectory $x^\mu = z^\mu (\tau)$ traced out
by the membrane in physical space.  One may think of the 
membrane as formed out of strings circling around the fourth and
fifth directions.  The parameters $g^4$ and $g^5$ 
in (\ref{current}) characterize the density of strings in
each direction. 
The above currents (being those of a particle)
are automatically conserved for any
choice of $g^4$ and $g^5$.
The field $G_{\Lambda \Gamma \Omega}$ has 
support on a four-dimensional surface
which may be taken to be a string worldsheet $y^{\lambda}
(\tau, \sigma)$ ending on the worldline $z^\mu(\tau)$
in $R^4$ times the torus (which we take to have unit 
volume).  Its non-vanishing components
do not depend on $x^4$ and $x^5$ and read
\begin{eqnarray}
G^{4 \lambda \mu} &=& - g^4 \int_{WS} 
\delta^{4}(x^\mu -y^ \mu)
dy^\lambda \wedge dy^\mu, \\
G^{5 \lambda \mu} &=& - g^5 \int_{WS}  
\delta^{4}(x^\mu -y^ \mu)
dy^\lambda \wedge dy^\mu.
\end{eqnarray}

Now, if we assume that $A_{\Lambda\Omega}$ does not
depend on the torus coordinates $x^4$ and $x^5$, the action
(\ref{actionsources2}) becomes a four-dimensional integral
since the integrand does not depend on the internal 
coordinates.
Assuming furthermore that the only non-vanishing
components of $A_{\Lambda\Omega}$ are
 $A_{4 i}$ and $A_{5 i}$
and making the identifications
\begin{eqnarray}
A_{4 i} &=& A^1_i, \;\;  A_{5 i} = -  A^2_i, \\
g^4 &=& q^2, \;\; g^5 = q^1
\end{eqnarray}
one finds that (\ref{actionsources2}) reduces exactly
to the manifestly duality-invariant action
(\ref{actionsources1}) for electromagnetism, once one has
of course added or reduced to $D=4$ the appropriate 
kinetic term for the source. From 
this dimensional reduction perspective, duality 
appears as a
spacetime transformation (rotation in the ``internal" 
4-5 space).
That duality in four dimensions can be obtained from the
chiral 2-form in six dimensions upon appropriate dimensional
reduction was previously observed in the sourceless case 
\cite{Verlinde,Berman,GNair}.

A more detailed report of the present work will be presented 
elsewhere \cite{future}.  There, we will also investigate
duality in the more general context of non-linear electrodynamics 
and show how the covariant equations determining duality-invariant 
theories \cite{GibbR}
arise in the Hamiltonian formalism through the Dirac-Schwinger
Lorentz invariance criterion on the commutator of the equal-time
Hamiltonian densities.

\section*{Acknowledgements}
The work of SD was supported by the National Science Foundation,
grant \#PHY-9315811, that of MH was partly supported
by a research grant from FNRS (Belgium),
that of AG and CT by Grants 3960008 and 1940203 
of FONDECYT (Chile).  AG and
CT also acknowledge institutional 
support to the
Centro de Estudios Cient\'{\i}ficos de Santiago provided by
SAREC (Sweden) and a group of Chilean private companies
(EMPRESAS CMPC, CGE, COPEC, MINERA LA ESCONDIDA, NOVAGAS
Transportandores de Chile, ENERSIS, BUSINESS DESIGN ASS., XEROX
Chile).


\begin{thebibliography}{100}

\bibitem{Olive} For a recent review, see D. I. Olive,
{\it Exact Electromagnetic Duality}, hep-th/9508089.

\bibitem{DHT} S. Deser and C. Teitelboim, {\em Phys. Rev.} {\bf D 13}
(1976) 1572; 
S. Deser, M. Henneaux and C. Teitelboim, {\em Phys. Rev.} {\bf  D 55}
(1997) 268.

\bibitem{Des} Some of the free field results
given above were recently summarized in S. Deser,
{\it Black Hole Electromagnetic Duality}, hep-th/9701157.

\bibitem{SSen} J.H. Schwarz and A. Sen, {\em Nucl. Phys.} {\bf B 411}
(1994) 35.

\bibitem{Berk} N. Berkovits, 
{\em Phys. Lett.} {\bf B 388} (1996) 743;  {\it Local
Actions with Electric and Magnetic Sources}, hep-th/9610134.

\bibitem{FloJa} R. Floreanini and R. Jackiw, {\em Phys.
Rev. Lett.} {\bf 59} (1987) 1873.

\bibitem{HT2} M. Henneaux and C. Teitelboim, {\em Phys.
Lett.}
{\bf 206 B} (1988) 650.

\bibitem{Dirac} P.A.M. Dirac, {\em Proc. Roy. Soc. London}
{\bf A 133} (1931) 60; {\em Phys. Rev.} {\bf 74} (1948) 817.

\bibitem{Schwinger} J. Schwinger, {\em Phys. Rev.} {\bf 173} 
(1968) 1536;
{\em Science} {\bf 165} (1969) 757.

\bibitem{Zwan} D. Zwanziger, {\em Phys. Rev.} {\bf 176} (1968) 1489.

\bibitem{Wu} T. T. Wu and C. N. Yang, {\em Phys. Rev. } {\bf D 12}
(1975) 3845; {\em Nucl. Phys.} {\bf B 107} (1976) 365.

\bibitem{DH} F. P. Devecchi and M. Henneaux, {\em Phys. Rev.}
{\bf D 54} (1996) 1606.

\bibitem{Bengtsson} I. Bengtsson and A. Kleppe, {\it
On Chiral P-Forms}, hep-th/9609102.

\bibitem{Pasti} P. Pasti, D. Sorokin and M. Tonin,  {\it On
Lorentz Invariant Actions for Chiral P-Forms},
hep-th/9611100.

\bibitem{CT} C. Teitelboim, {\em Phys. Lett} {\bf B 167} 
(1986) 63, 67.

\bibitem{HT}  M. Henneaux and C. Teitelboim, {\em Found. Phys.}
{\bf 16} (1986) 593.

\bibitem{Nepo} R. Nepomechie {\em Phys. Rev.} {\bf D 31} (1985) 
1921.

\bibitem{Perry} M. Perry and J.H. Schwarz, {\it Interacting Chiral
Gauge Fields in Six Dimensions and Born-Infeld Theory},
hep-th/9611065.

\bibitem{Verlinde} E. Verlinde, {\em Nucl. Phys.} {\bf B 455}
(1995) 211.

\bibitem{Berman} D. Berman, {\it Classical Duality from
Compactification of Self-Dual Five Form Maxwell
Theory in Ten Dimensions}, hep-th/9612191.

\bibitem{GNair} I. Giannakis and V. P. Nair, {\it Symplectic
Structures and Self-Dual Fields in
$(4k+2)$ Dimensions}, hep-th/9702024.

\bibitem{future} S. Deser, A. Gomberoff, M. Henneaux,  \"{O}.
Sar{\i}o\u{g}lu  and C. Teitelboim, {\em in preparation}.

\bibitem{GibbR} G. W. Gibbons and D. A. Rasheed,
{\em Nucl. Phys.}
{\bf B 454} (1995) 185; I. Bengtsson, {\it Manifest Duality in
Born-Infeld Theory}, hep-th/9612174.

\end{thebibliography}
\end{document}